\documentclass[showpacs,twocolumn,aps]{revtex4}
\usepackage{amssymb}
\usepackage{amsmath}
\usepackage{graphicx}
\usepackage{lscape}
\begin{document}
\title{Parton rescattering effect on the charged hadron forward-backward
multiplicity  correlation in $pp$ collisions at $\sqrt s$=200 GeV}
\author{Yu-Liang Yan$^{1}$, Bao-Guo Dong$^{1,2}$, Dai-Mei Zhou$^{3}$, Xiao-Mei
Li$^{1}$, Hai-Liang Ma$^{1}$, Ben-Hao Sa$^{1,3,4}$\footnote{Corresponding
author: sabh@ciae.ac.cn}}
\address{ 1 China Institute of Atomic Energy, P.O. Box 275(18), Beijing
102413, China \\
2 Center of Theoretical Nuclear Physics, National Laboratory of
Heavy Ion Collisions, Lanzhou 730000, China\\
3 Institute of Particle Physics, Huazhong Normal University, Wuhan
430079, China\\
4 CCAST (World Laboratory), P. O. Box 8730 Beijing 100080, China}

\begin{abstract}
The parton rescattering effect on the charged hadron
forward-backward multiplicity correlation in $pp$ collisions at
$\sqrt s$=200 GeV is studied by a parton and hadron cascade model,
PACIAE, based on the PYTHIA model. The calculated multiplicity and
pseudorapidity distribution of the final state charged hadron are
well compared with experimental data. It turned out that the final
state charged hadron pseudorapidity distribution are different from
the initial state charged partons. The parton rescattering effect on
the charged hadron forward-backward multiplicity correlation
increases with increasing parton rescattering strength in the center
pseudorapidity region ($\vert\eta\vert <1$). However, this effect
becomes weaker in the outer pseudorapidity region ($\vert\eta\vert>
1$).

\end{abstract}

\maketitle

\section {INTRODUCTION}
The study of fluctuations and correlations has been suggested as a
useful means to reveal the mechanism of particle production and the
formation of Quark-Gluon-Plasma (QGP) in the relativistic heavy-ion
collisions \cite{hwa2,naya}. Correlations and fluctuations of the
thermodynamic quantities and/or the produced particle distributions
could be significantly altered when the system undergoes phase
transition because the degrees of freedom is largely different
between the hadronic matter and the quark-gluon matter.

The experimental study of fluctuations and correlations becomes a
hot topic in relativistic heavy ion collisions with the availability
of high multiplicity event-by-event measurements at the CERN-SPS and
BNL-RHIC experiments. An abundant experimental data have been
reported \cite{appe,afan,star,star1,star2,phen,phen1,phen2,phob}
where a lot of new physics are arisen and urgent to be studied.

A lot of theoretical investigations have also been reported such as
Refs. \cite{dpm,meng,paja,hwa1,yan}. We have used the PYTHIA model
to investigate the strength of charged hadron forward-backward
multiplicity correlation in $\bar pp$ and $pp$ collisions at $\sqrt
s$=200 GeV \cite{yan}. It argued that a factor of 3-4 apparent
discrepancy between UA5 $\bar pp$ data \cite{ua5} and STAR $pp$ data
\cite{star4} can be attributed to the differences in detector
acceptances and observed bin interval in both experiments. In
\cite{hwa1} the back-to-back parton scattering was considered as the
origin of final state hadronic covariance. They assumed the details
of hadronization are not essential and related the back-to-back parton
scattering angles to a Gaussian-like hadronization function. Then
they derived the final state charged hadron forward-backward multiplicity
covariance. Without more dynamical inputs, their results
were well compared with STAR data \cite{star4}, ``thus dispelling
the notion that correlation length has any fundamental significance".
Stimulated by this interesting conclusion, a parton and hadron
transport model, PACIAE \cite{sa}, is employed in this paper to
investigate the effect of parton rescattering in parton evolution
stage on the final state charged hadron forward-backward multiplicity
correlation in the $pp$ collisions at $\sqrt s$= 200 GeV.

Following \cite{cape} the strength of charged particle forward-backward
multiplicity correlation, $b$, is defined as
\begin{equation}
\ b =\frac{\langle n_fn_b\rangle - \langle n_f\rangle \langle
n_b\rangle}{\langle n_f^2\rangle - \langle n_f\rangle^2} =
\frac{D_{fb}^2}{D_{ff}^2},
\end{equation}
where $n_f$ and $n_b$ are, respectively, the number of charged
particles in forward and backward pseudorapidity bins defined
relatively and symmetrically to a given pseudorapidity $\eta$. The
$\langle n_f\rangle$ ($\langle n_b \rangle$) is the mean value of
$n_f$ ($n_b$), the $D_{fb}^2$ and $D_{ff}^2$ are the
forward-backward multiplicity covariance and forward multiplicity
variance, respectively. One always studies $b$ as a function of the
center distance of two forward and backward pseudorapidity bins
($\Delta\eta$) and the $\eta$ acceptance.

\section {THE PACIAE MODEL}
The parton and hadron cascade model, PACIAE \cite{sa}, is based on
PYTHIA \cite{soj2} which is a model for high energy hadron-hadron ($hh$)
collisions. The PACIAE model is composed of four stages: parton
initialization, parton evolution (rescattering), hadronization, and
hadron evolution (rescattering).

1. PARTON INITIALIZATION: In the PACIAE model, a nucleus-nucleus
collision is decomposed into the nucleon-nucleon ($NN$) collisions
based on the collision geometry. The $NN$ collision is described
with the PYTHIA model, i.e. it is decomposed into the parton-parton
collisions. The hard parton-parton collision is described by the
lowest-leading-order (LO) pQCD parton-parton cross section
\cite{comb} with modification of parton distribution function in the
nucleon. And the soft parton-parton interaction is considered
empirically. The semihard, between hard and soft, QCD $2\rightarrow
2$ processes are also involved in the PYTHIA (PACIAE) model. Because
the initial- and final-state QCD radiation added to the
parton-parton collision process, the PYTHIA (PACIAE) model generates
a partonic multijet event composed of quark pairs, diquark pairs and
gluons for a $NN$ ($hh$) collision. That is followed, in the PYTHIA
model, by the string-based fragmentation scheme (Lund string model
and/or Independent Fragmentation model). Thus a hadronic final state
is reached for a $NN$ ($hh$) collision. However, in the PACIAE model
above fragmentation is switched off temporarily, so the result is a
partonic multijet event instead of a hadronic state. If the diquarks
(anti-diquarks) are split forcibly into quarks (anti-quarks)
randomly, the consequence of a $NN$ ($hh$) collision, thus a
nucleus-nucleus collision, is its initial partonic state composed of
quarks, anti-quarks, and gluons.

2. PARTON EVOLUTION: The next stage in the PACIAE model is the
parton evolution (parton rescattering). Here the $2\rightarrow 2$
LO-pQCD differential cross sections \cite{comb} are employed. The
differential cross section of a subprocess $ij\rightarrow kl$ reads
\begin{equation}
\frac{d\sigma_{ij\rightarrow
kl}}{d\hat{t}}=K\frac{\pi\alpha_s^2}{\hat{s}}\sum_{ij\rightarrow kl},
\end{equation}
where the $K$ factor is introduced for considering the higher order pQCD
and non-perturbative QCD corrections as usual and $\alpha_s$ stands for the
effective strong coupling constant. Taking the process $q_1q_2 \rightarrow
q_1q_2$ as an example one has
\begin{equation}
\sum_{q_1q_2\rightarrow
q_1q_2}=\frac{4}{9}\frac{\hat{s}^2+\hat{u}^2}{\hat{t}^2},
\label{eq3}
\end{equation}
where the $\hat{s}$, $\hat{t}$, and $\hat{u}$ are the Mandelstam
variables. Since it diverges at $\hat{t}$=0, it has to be
regularized by introducing the parton colour screen mass $\mu$ as
follows
\begin{equation}
\sum_{q_1q_2\rightarrow
q_1q_2}=\frac{4}{9}\frac{\hat{s}^2+\hat{u}^2}{(\hat{t}-\mu^2)^2}.
\end{equation}

The total cross section of the parton collision $i+j$ then reads
\begin{equation}
\sigma_{ij}(\hat{s})=\sum_{k,l}\int_{-\hat{s}}^{0}d\hat{t}
\frac{d\sigma_{ij\to kl}}{d\hat{t}}.
\end{equation}
With above total and differential cross sections the parton
evolution (parton rescattering) can be simulated by the Monte Carlo
method.

3. HADRONIZATION: The parton evolution stage is followed by the
hadronization at the moment of partonic freeze-out (no more parton
collision at all). In the PACIAE model, the phenomenological
fragmentation model and coalescence model are supplied for the
hadronization of partons after rescattering. The String
Fragmentation (SF) model is adopted in this paper. We refer to
\cite{sa} for the details of the hadronization stage.

4. HADRON EVOLUTION: After hadronization the rescattering among
produced hadrons is dealt with the usual two-body collision model.
We neglect the hadronic rescattering in $pp$ collisions as usual.
The details of hadronic rescattering see \cite{sa1}.

\begin{table}[htbp]
\caption{Charged particle multiplicity ($|\eta|\leq 5.4$) in $pp$
collisions at $\sqrt{s}$=200 GeV.}
\begin{tabular}{clll}
\hline\hline
                    & \multicolumn{2}{c} {PACIAE}   \\
\cline{2-3}
                    &  parton$^1$ & hadron$^2$ & \raisebox{2ex}[2ex]
                       {Exp. data$^{2,3}$ }\\
\hline
  no parton scat. & 5.96  &21.2 (17.7)$^4$&                   \\
 18\% parton scat. & 6.13  &21.7 (18.5)    &19.9 $\pm$ 2.2    \\
 58\% parton scat.& 6.15  &21.9 (18.7)    &                    \\

\hline\hline \multicolumn{4}{l}{$^1$ after parton rescattering, in
full $\eta$ phase
                   space.}\\
\multicolumn{4}{l}{$^2$ in final hadronic state.}\\
\multicolumn{4}{l}{$^3$ inelastic $pp$ collision data taken from
                   \cite{phob1}.}\\
\multicolumn{4}{l}{$^4$ value given in bracket is calculated for
inelastic $pp$ collisions.}
\end{tabular}
\label{mul}
\end{table}

\section {PARTON RESCATTERING EFFECT ON HADRON MULTIPLICITY CORRELATION}
As we aim at the physics behind the experimental data rather than
reproducing the data, model parameters are fixed in the calculations.
All calculations are for $\sqrt{s} $=200 GeV Non-Single Diffractive (NSD)
collisions except that marked especially. In order to see the parton
rescattering effect in the parton evolution stage, we design no, weak,
and strong parton rescattering cases as follows:
\begin{itemize}
\item No parton rescattering: $K$=0, without parton rescattering at all.
\item Weak parton rescattering: $K$=1 and $\mu^2$=0.4 GeV$^2$/c$^4$, with
      nearly 18\% charged partons participating the rescattering.
\item Strong parton rescattering: $K$=3 and $\mu^2$=0.1 GeV$^2$/c$^4$,
      with nearly 58\% charged partons participating the rescattering.
\end{itemize}

\begin{figure}[htbp]
\includegraphics[width=3.0in,angle=0]{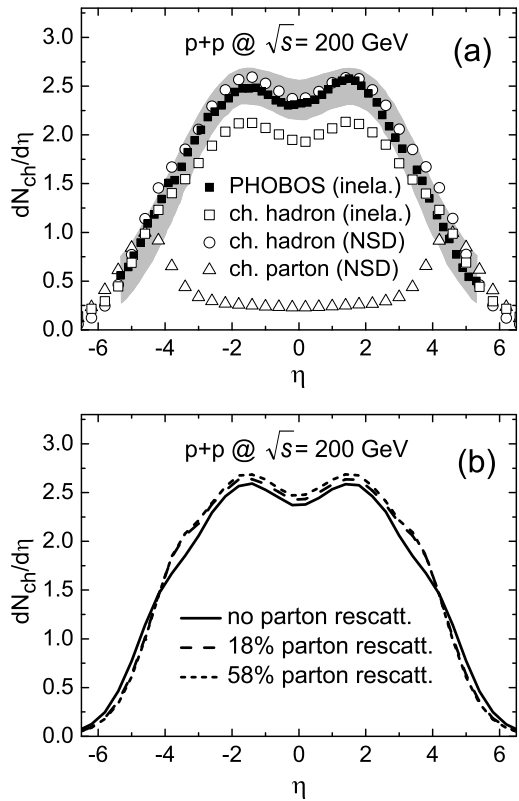}
\vspace{-0.1in} \caption{(a) The pseudorapidity distributions of the
initial state charged partons and final state charged hadrons and
(b) the parton rescattering effect on the final state charged hadron
pseudorapidity distribution in $pp$ collisions at $\sqrt{s}$= 200
GeV. The PHOBOS inelastic $\eta$ distribution data are taken from
\cite{phob1}.} \label{fig1}
\end{figure}

\begin{figure}[htbp]
\includegraphics[width=3.2in,angle=0]{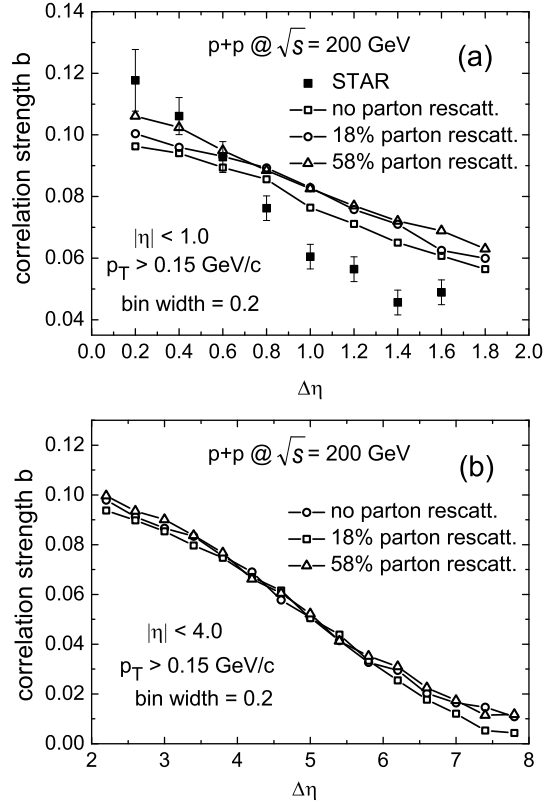}
\vspace{-0.1in} \caption{The parton rescattering effect on the
final state charged hadron forward-backward multiplicity correlation strength
in (a) central pseudorapidity range ($|\eta|<1$)
and (b) outer pseudorapidity range ($1<|\eta|<4$) in $pp$ collisions
at $\sqrt{s}$=200 GeV. The open squares, open circles and open triangles
are for case of no, 18\% and 58\% participant parton, respectively.
The experimental data are taken from \cite{star4}.} \label{fig2}
\end{figure}

Table \ref{mul} shows that although we did not adjust the model
parameters the PHOBOS multiplicity data of final state charged
hadrons\cite{phob1}  are well reproduced. From the PACIAE results in
this table one sees that the multiplicity of initial state charged
partons ($u+d+s$ and their anti-particles) and final state charged
hadrons increase weakly from no, to weak, and to strong parton
rescattering case. This increase in percentage is not as strong as
the increase in percentage of participating charged parton given in
the rescattering case definition. This is because the total number
of initial state charged partons is just around 6 and only the
$2\rightarrow 2$ parton rescattering processes are considered in the
parton evolution stage.

\begin{figure}[htbp]
\includegraphics[width=3.5in,angle=0]{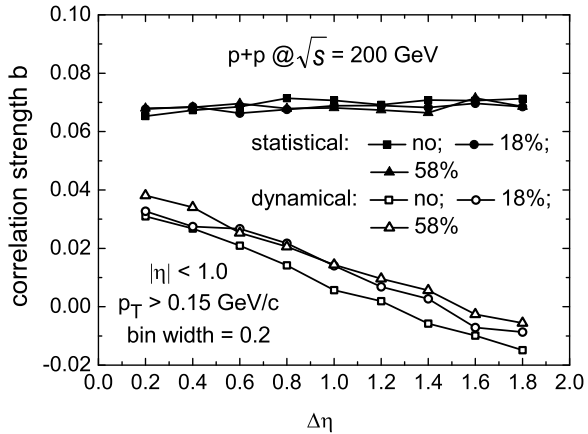}
\vspace{-0.1in} \caption{The parton rescattering effect on the
statistical and dynamical correlations of final state charged
hadron in central pseudorapidity range ($|\eta|<1$) in $pp$
collisions at $\sqrt{s}$=200 GeV.} \label{fig3}
\end{figure}

The calculated pseudorapidity distributions of initial state charged
partons and final state charged hadrons are compared in
Fig.~\ref{fig1}(a). The full and open squares are, respectively, the
PHOBOS inelastic $pp$ collision data (taken from \cite{phob1}) and
the corresponding theoretical results for final state charged hadrons.
And the open circles are theoretical results calculated for final state
charged hadrons in NSD $pp$ collisions. One sees that the PHOBOS data
agree well with the NSD calculations but are higher than the inelastic
calculations in the center $\eta$ region. This is consistent with the
fact that the charged hadron multiplicity in inelastic calculations is
lower than the PHOBOS data as shown in Tab. \ref{mul}. Comparing the
$\eta$ distribution of final state charged hadrons to the one of initial
state charged partons (open triangles) we know that the
parton rescattering and fragmentation fill up the wide valley between
two fragmentation peaks at $|\eta| \sim 5$ in the $\eta$ distribution
of initial state charged partons.

Fig.~\ref{fig1}(b) shows the parton rescattering effect on the
pseudorapidity distribution of final state charged hadrons. One
sees that in the $\vert\eta \vert\leq4$ region the pseudorapidity
distributions move upward monotonously from no, to weak, and to
strong parton rescattering case. However, in the outer region one
sees nearly the opposed situation.

Figure~\ref{fig2} shows the parton rescattering effect on the
forward-backward multiplicity correlation strength of final state
charged hadron in $pp$ collisions at $\sqrt{s}$=200 GeV. The
corresponding experimental data \cite{star4} are given by full
squares. We see in panel (a) that the correlation strength increases
with increasing strength of parton rescattering in the central
pseudorapidity region of $\vert\eta\vert <1$ (or $\Delta \eta <2$).
While panel (b) shows that the parton rescattering effect becomes
weaker in the outer pseudorapidity region $|\eta|>1$ (or $\Delta
\eta >2$). It has to point out here that the $\eta$ acceptance in
panel (b) is nearly four times larger than the one in panel (a) so
the $b$ at $\Delta \eta =2$ in panel (b), for instance, is larger
than the corresponding one in panel (a).

We also use the mixed events method \cite{yan} to study the parton
rescattering effect on the statistical and dynamical
(non-statistical) correlations separately. The results are shown in
Fig.~\ref{fig3}. We see that the initial state parton rescattering
have almost no effect on statistical correlations. That is because
the statistical correlation is steaming from the multiplicity
fluctuation in the detected $\eta$ bins and the parton rescattering
have only little effect on the multiplicity of final state hadrons
as shown in Tab. \ref{mul}. However, one sees that the dynamical
correlation strength is increasing monotonously from no, to weak,
and to strong parton rescattering case in central pseudorapidity
range ($|\eta|<1$). That shows the parton rescattering increase the
dynamical correlation among partons, it survived the hadronization,
and effected on the dynamical correlation of final state hadrons.

\section {SUMMARY}
Using a parton and hadron cascade model PACIAE we have investigated
the parton rescattering effect on forward-backward multiplicity
correlation strength of the final state charged hadron in $pp$
collisions at $\sqrt s=200$ GeV. The calculated multiplicity and
pseudorapidity distributions of the final state charged hadron are
well compared with the corresponding experimental data. It turned
out that the pseudorapidity distribution of final state charged
hadron is different from the initial state charged parton. The
parton rescattering effect on the forward-backward multiplicity
correlation of the final state charged hadron increases with
increasing parton rescattering strength in the center pseudorapidity
region ($\vert\eta\vert < 1$). This increase becomes weaker in the
outer pseudorapidity region ($\vert\eta\vert> 1$). However, the
final state hadron correlation strength is related to the initial
state partons.

The parton rescattering effect on the forward-backward multiplicity
correlation of final state charged hadron is not as strong as
expected from the increasing of participating charged parton. That
is because the average total number of initial state charged partons
is just around 6 in above collisions and the parton rescatterings
considered in parton evolution stage are all $2\rightarrow 2$
processes. This is consistent with the notion that the small
probability of QGP formation in $pp$ collisions at RHIC. One may
expect much stronger parton rescattering effect in the
nucleus-nucleus collisions at the same energy and even in the $pp$
collisions at LHC energy.

The financial support from NSFC (10635020, 10605040, 10705012, and
10875174) in China is acknowledged.


\begin{references}
\bibitem{hwa2} R. C. Hwa, arXiv:nucl-th/0701053v1.
\bibitem{naya}  T. K. Nayak, J. of Phys. G {\bf 32}, S187 (2006); arXiv:nucl-ex/0608021v1.
\bibitem{appe} H. Appelsh$\ddot{a}$user, et al., NA49 Collaboration, Phys.
 Lett. B {\bf 459},  679 (1999).
\bibitem{afan} S. V. Afanasiev, et al., NA49 Collaboration, Phys. Rev. Lett.
 {\bf 86}, 1965 (2001).
\bibitem{star}  J. Adams, et al., STAR Collaboration, Phys. Rev. C {\bf68},
 044905 (2003).
\bibitem{star1} J. Adams, et al., STAR Collaboration, J. Phys. G {\bf 32}, L37
(2006).
\bibitem{star2} J. Adams, et al., STAR Collaboration, Phys. Rev. C {\bf75}, 034901
(2007).
\bibitem{phen} K. Adcox, et al., PHENIX Collaboration, Phys. Rev. Lett.
 {\bf 89}, 082301 (2002).
\bibitem{phen1} S. S. Adler, et al., PHENIX Collaboration, Phys. Rev. Lett.
 {\bf 93}, 092301 (2004).
\bibitem{phen2} A. Adare, et al., PHENIX Collaboration, Phys. Rev. Lett.
 {\bf 98}, 232302 (2007).
\bibitem{phob} Zheng-Wei Chai, et al., PHOBOS Collaboration, J. of Phys.:
 Conference Series {\bf 27}, 128 (2005).
\bibitem{dpm} A. Capella and J. Tran Thanh Van, Z. Phys, C {\bf 18}, 85 (1983).
\bibitem{meng} Lian-sou Liu and Ta-chung Meng, Phys. Rev. D {\bf 27}, 2640 (1983),
 ibid D {\bf 33}, 1287 (1986).
\bibitem{paja}
N. S. Amelin, N. Armesto, M. A. Braun, E. G. Ferreiro, and C.
Pajares, Phys. Rev. Lett. {\bf 73}, 2813 (1994); N. Armesto, M. A.
Braun, and C. Pajares, Phys. Rev. C {\bf75}, 054902 (2007).
\bibitem{hwa1} R. C. Hwa and C. B. Yang , arXiv:0705.3073v1[nucl-th].
\bibitem{yan} Yu-Liang Yan, Bao-Guo Dong, Dai-Mei Zhou, Xiao-Mei Li,
and Ben-Hao Sa, Phys. Lett. B {\bf 660}, 478 (2008);
arXiv:0710.2187v2[nucl-th].
\bibitem{ua5} R. E. Ansorge et al., UA5 Collaboration, Z. Phys. C {\bf 37}, 191
(1988).
\bibitem{star4} B. K. Srivastavs, STAR Collaboration, Int. J. Mod. Phys. E {\bf 16}, 3363 (2008).
\bibitem{sa} Ben-Hao Sa, Xiao-Mei Li, Shou-Yang Hu, Shou-Ping Li, Jing Feng,
and Dai-Mei Zhou, Phys. Rev. C {\bf75}, 054912 (2007); Yu-Liang Yan,
Dai-Mei Zhou, Bao-Guo Dong, Xiao-Mei Li, Hai-Liang Ma, and Ben-Hao
Sa, Phys. Rev. C {\bf 79}, 054902 (2009);
arXiv:0903.0915v2[nucl-th].
\bibitem{cape} A. Capella, U. Sukhatme, C.-I. Tan, and J. Tran Thanh Van, Phys. Rep.
{\bf 236}, 225 (1994).
\bibitem{soj2} T. S\"ojstrand, S. Mrenna, and P. Skands, J. High Energy Phys.
 {\bf JHEP05}, 026 (2006); arXiv:hep-ph/0603175v1.
\bibitem{comb} B. L. Combridge, J. Kripfgang, and J. Ranft, Phys. Lett. B
 {\bf 70}, 234 (1977).
\bibitem{sa1} Ben-Hao Sa and Tai An, Comput. Phys. Commun. {\bf 90}, 121 (1995);
Tai An and Ben-Hao Sa, Comput. Phys. Commun. {\bf 116}, 353 (1999).
\bibitem{phob1} R. Nouicer, et al., PHOBOS Collaboration, J. Phys. G {\bf 30}, S1133
(2004); arXiv:nucl-ex/0403033v1.
\end{references}
\end{document}